\documentclass[fleqn,12pt]{wlscirep}

\usepackage{graphicx}
\usepackage{dcolumn}
\usepackage{bm}
\usepackage{CJK}
\usepackage{amsfonts}
\usepackage{amsmath}
\usepackage{amssymb}
\usepackage{txfonts}%
\usepackage{color}

\title{6.2-GHz modulated terahertz light detection using fast terahertz quantum well photodetectors}

\author[1,*]{Hua Li}
\author[1]{Wen-Jian Wan}
\author[1]{Zhi-Yong Tan}
\author[1]{Zhang-Long Fu}
\author[1]{Hai-Xia Wang}
\author[1]{Tao Zhou}
\author[1]{Zi-Ping Li}
\author[1]{Chang Wang}
\author[2]{Xu-Guang Guo}
\author[1,*]{Jun-Cheng Cao}

\affil[1]{Key Laboratory of Terahertz Solid-State Technology, Shanghai Institute of Microsystem and Information Technology, Chinese Academy of Sciences, 865 Changning Road, Shanghai 200050, China.}
\affil[2]{School of Optical-Electrical and Computer Engineering, University of Shanghai for Science and Technology, 516 Jungong road, Shanghai 200093, China.}

\affil[*]{hua.li@mail.sim.ac.cn and jccao@mail.sim.ac.cn}


\begin{abstract}

The fast detection of terahertz radiation is of great importance for various applications such as fast imaging, high speed communications, and spectroscopy. Most commercial products capable of sensitively responding the terahertz radiation are thermal detectors, i.e., pyroelectric sensors and bolometers. This class of terahertz detectors is normally characterized by low modulation frequency (dozens or hundreds of Hz). Here we demonstrate the first fast semiconductor-based terahertz quantum well photodetectors by carefully designing the device structure and microwave transmission line for high frequency signal extraction. Modulation response bandwidth of gigahertz level is obtained. As an example, the 6.2-GHz modulated terahertz light emitted from a Fabry-P\'{e}rot terahertz quantum cascade laser is successfully detected using the fast terahertz quantum well photodetector. In addition to the fast terahertz detection, the technique presented in this work can also facilitate the frequency stability or phase noise characterizations for terahertz quantum cascade lasers.

\end{abstract}
\begin{document}

\flushbottom
\maketitle
%
%
The terahertz wave \cite{THz} with the frequency defined between 0.1 and 10 THz is of great interest due to its unique properties such as transparency for papers and plastic materials, abundant absorption ``fingerprints" of various chemicals, and potential large communication bandwidth.
In the previous dozens of years, the fast development of terahertz radiation sources (such as quantum cascade lasers (QCLs) \cite{1stTHzQCL,THzQCLReview}, uni-travelling-carrier photodiodes \cite{UTCPD}, frequency multipliers \cite{FMulti}, etc.) and detectors (pyroelectrics \cite{pyro}, bolometers \cite{Bolometer}, Golay-cell \cite{Golay}, etc.) results in a significant progress of terahertz technology. Laboratory applications for example the terahertz imaging, data transmission, and spectroscopy have been already demonstrated \cite{3Dimaging,ZhouImaging,ChenEL,Yang,FuSR}. Further advances in terahertz technology require fast detection of gigahertz frequency modulated terahertz radiation. The fast terahertz detector is the last component for implementing fast terahertz applications since the terahertz source employing the quantum cascade design has been already verified for being fast modulated upto 35 GHz \cite{GellieOE}. Regarding the terahertz detectors, most thermal infrared detectors working at low modulation frequencies (dozens or hundreds of Hz) are therefore not able to detect the fast modulated terahertz wave. However, the quantum well photodectors (QWPs) \cite{LiuQWP,Guo2013} employing the electron intersubband transitions are supposed to work fast in the terahertz region due to the fast carrier relaxation time in picoseconds level. Actually in the mid-infrared wavelength range, room temperature high frequency heterodyne detection upto 110 GHz with small mesa quantum well infrared photodetectors has been already demonstrated in 2006 employing the air bridge technique and coplanar waveguide configuration \cite{Grant2006}. However, in the terahertz regime, due to the diffraction limit the terahertz light coupling to a small QWP device is relatively difficult. And therefore, the high microwave frequency operation of terahertz QWP has been not yet reported.

Here in this work by utilizing a carefully designed microwave transmission line to efficiently extract the radio frequency (RF) signal, we first demonstrate the fast detection of GHz-modulated terahertz radiation using a terahertz QWP with a device area of 400$\times$400 $\mu$m$^2$. The electrical rectification measurements reveal that the QWP device has a modulation response bandwidth upto 4.3 GHz. As a proof for the fast detection, a 6.2-GHz modulated terahertz light emitted from a long cavity terahertz QCL was successfully detected in terms of inter-mode optical beat note spectrum using the fast terahertz QWP.

\section*{Experimental setup}

\begin{figure}[h]
\centering
\includegraphics[width=0.7\textwidth]{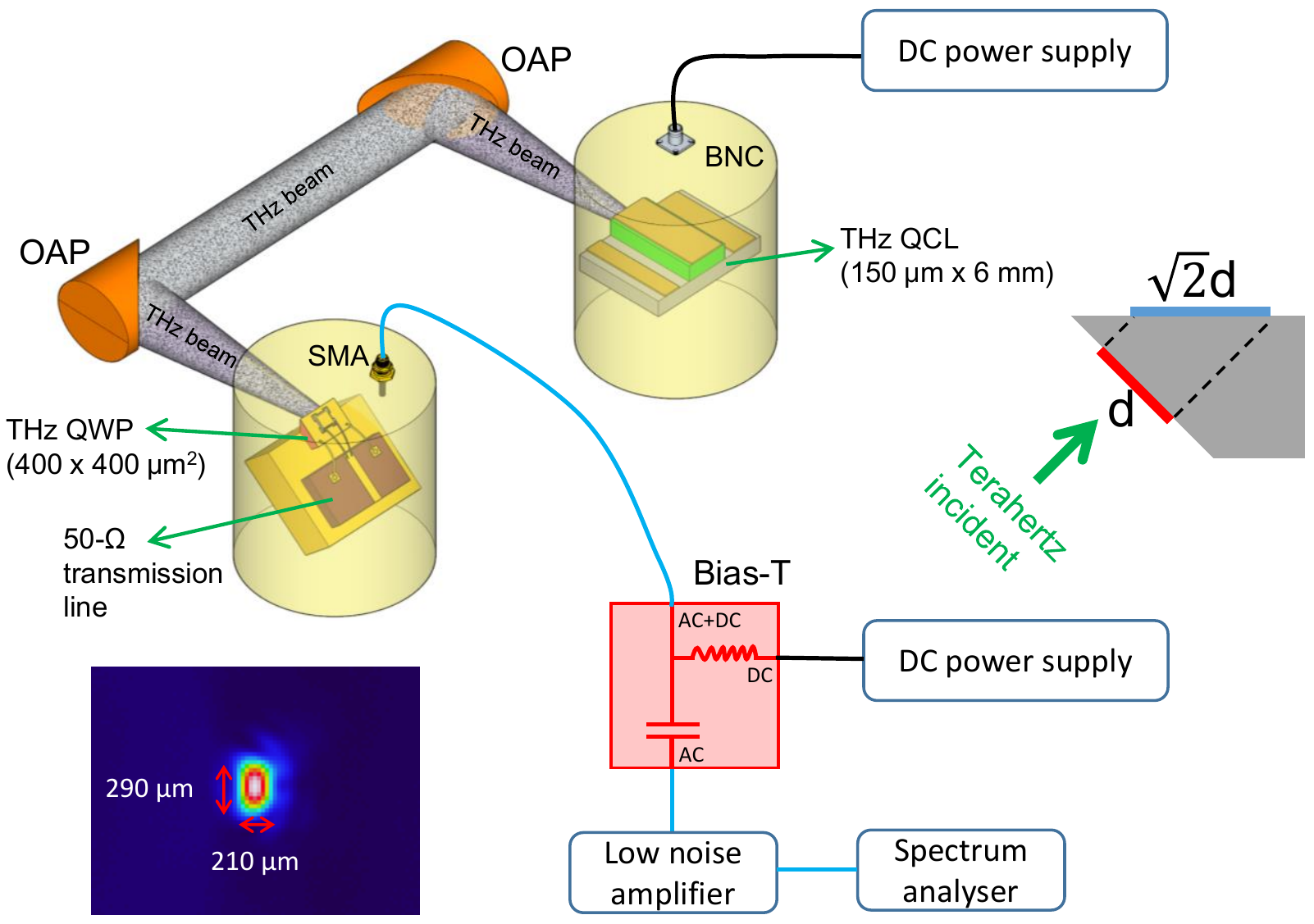}
\caption{\textbf{Experimental setup of the fast terahertz detection using a microwave transmission line equipped terahertz QWP.} The modulated terahertz light is generated from a 6-mm long cavity terahertz QCL working in continuous wave mode. Two off-axis parabolic (OAP) mirrors are used for light collection and collimation. The lower left inset shows the two-dimensional focused beam pattern of the terahertz QCL measured using a terahertz imager. The upper right inset is the schematic of the terahertz light incident on the QWP chip. $d$ is the diameter of the terahertz beam and $\sqrt{2}$$d$ the side dimension of the QWP chip to ideally absorb the terahertz radiation. SMA: SubMiniature version A, BNC: Bayonet Neil Concelman.}
\label{setup}
\end{figure}

Figure \ref{setup} shows the experimental setup employed for the fast terahertz detection. The terahertz light is generated from an electrically-pump terahertz QCL. Due to the resonant cavity, the laser light is inherently modulated by the cavity round trip frequency of $\sim$$c/2ln$ with $c$ being the speed of light, $l$ the cavity length, and $n$ the refractive index. Two parabolic mirrors are used to collect and focus the terahertz light onto the QWP. To couple the polarized terahertz light into the QWP mesa, a 45$^\circ$ edge facet geometry is used. To facilitate the extraction of the high frequency signal from the terahertz QWP mesa, a 50-$\Omega$ microwave transmission line is mounted as close as possible to the QWP mesa as shown in Fig. \ref{setup}. The transmission line consists of an AlN dielectric substrate which is gold-coated on the backside and a thin metallic layer gold on top of the substrate. To efficiently collect the microwave signal via the transmission line, the design criteria is to make the dielectric layer the same thickness as the terahertz chip thickness around 200 $\mu$m and in the meantime set the transmission line impedance to 50-$\Omega$ by adapting the width of the metallic line. The top contact of the QWP mesa is wire bonded and connected to the central strip line while the back contact to the gold-plated holes which are connected to the ground. To reduce the microwave signal attenuation as much as possible and apply voltage onto the QWP mesa, high performance coaxial radio frequency cables (Huber+Suhner Sucoflex 100) and SMA connectors are used. A bias-T with a bandwidth of 40 GHz and DC current limit of 0.5 A is used to inject DC voltage to the QWP and simultaneously extract the high frequency signal via the AC port. The weak microwave signal is firstly amplified by a low noise amplifier with a gain of 30 dB and then sent to a spectrum analyser (upto 26.5 GHz) for characterizations.

The lower left inset of Fig. \ref{setup} shows the focused terahertz beam pattern measured by replacing the QWP with a terahertz imager. It can be seen that with two parabolic mirrors, the terahertz light can be focused in a quasi-Gaussian area at the focus of the second parabolic mirror. The vertical and horizontal widths of the beam pattern are measured to be 290 and 210 $\mu$m, respectively, whilst the wavelength of the laser emission is around 71 $\mu$m. The upper right inset figure shows the schematic geometry of the QWP chip as well as the incident terahertz wave. If we ideally assume the incident terahertz beam has a gaussian pattern with a diameter of $d=300$ $\mu$m, the side length of the square mesa is supposed to be $\sqrt{2}d\approx420$ $\mu$m to fully absorb the incident terahertz emission. Because of this, we finally choose the mesa area of 400$\times$400 $\mu$m$^2$ in this work to balance the trade-off between photoresponse and device capacitance.

\section*{Results}

\begin{figure}[t]
\centering
\includegraphics[width=0.65\textwidth]{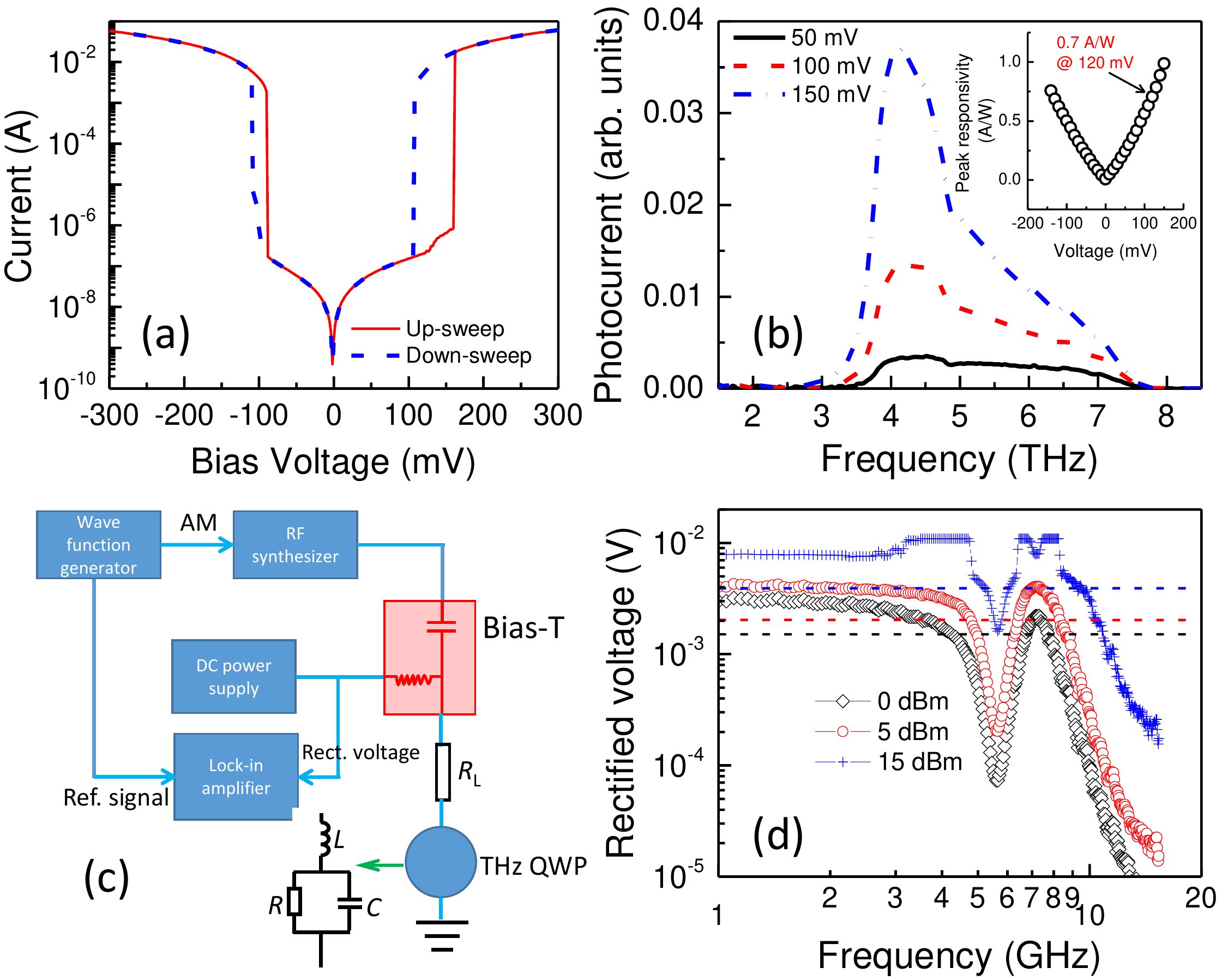}
\caption{\textbf{Device characterizations of the terahertz QWP.} (a) The dark current versus voltage characteristics of the 400$\times$400 $\mu$m$^2$ terahertz QWP measured at 5 K. (b) Photoresponse spectra of the QWP device at different bias voltages. The inset plots the measured peak responsivity as a function of voltage. (c) Experimental setup of the rectification measurement. $R_L$ represents the 50-$\Omega$ transmission line. The inset is the equivalent circuit of the QWP device for microwave modelling. (d) The rectifications at different microwave power as the QWP is biased at 120 mV. The dashed lines show the -3 dB levels for 0, 5, and 15 dBm power values.}
\label{QWP}
\end{figure}

Before demonstrating the fast terahertz detection, we first characterize the electrical and optical performances of the terahertz QWP. The measured dark current versus voltage characteristics of the 400$\times$400 $\mu$m$^2$ terahertz QWP at 5 K is shown in Fig. \ref{QWP}a, where the electrical hysteresis can be clearly observed. The QWP device is working at low current level of 10$^{-7}$ A below 200 mV to achieve moderate background noise. Figure \ref{QWP}b plots the photoresponse spectra of the QWP at different bias voltages measured using a fast-scan Fourier Transform Infrared Spectroscopy. It can be seen that the spectra are peaked around 4.2 THz. The inset of Fig. \ref{QWP}b gives the peak responsivity as a function of voltage calibrated  with a blackbody source. In the working voltage range between 100 and 150 mV, the QWP device shows a peak responsivity of 0.7 A/W. By measuring the noise spectral density expressed in A/$\sqrt{\textrm{Hz}}$, we finally obtain the noise equivalent power (NEP) of 0.4 pW/$\sqrt{\textrm{Hz}}$. Both the responsivity and NEP parameters show that the QWP is a sensitive detector for terahertz emission.

Due to the complexity or the difficulty of performing optical heterodyne measurements in the teraherz regime, the microwave rectification technique \cite{LiuJQE1996} is an useful and simple way to investigate the high frequency characteristics of the QWP device. Instead of using two terahertz sources with a frequency separation in microwave range for heterodyne measurements, as shown in Fig. \ref{QWP}c we apply a microwave signal to the QWP via a Bias-T and measure the change in the DC voltage. The rectification relies on the nonlinearity of the voltage-current (\textit{V-I}) curve and the microwave modulation amplitude. The rectified DC voltage at current $I_0$, $V_{\textrm{rect}}$, can be written as \cite{13GHz,HinkovOE}
\begin{equation}\label{rectV}
V_{\textrm{rect}}=\frac{1}{2}|V^{\prime\prime}|_{I_0}I_{\textrm{RF,QWP}}^2,
\end{equation}
where $|V^{\prime\prime}|_{I_0}$ is the second derivative of the $V-I$ curve at $I_0$ and $I_{\textrm{RF,QWP}}$ is the effective current amplitude applied by the radio frequency (RF) source to the QWP device. The inset gives the parallel resistance-capacitance (R-C) equivalent circuit of the QWP device for modeling that will be discussed in detail later. The inductance $L$ is introduced by the wire bonds used for connecting the QWP mesa and the microwave transmission line. $R$ and $C$ are resistance and capacitance of the QWP mesa, respectively.

Figure \ref{QWP}d shows the measured rectified voltage as a function of frequency for the QWP device at various injecting microwave power. The bias voltage is set to an optimal value of 120 mV which corresponds to an electric field of 0.41 kV/cm. The measured 3 dB response bandwidth $f_{3 \textrm{dB}}$ increases from 4.3 to 5.3 GHz with increasing the microwave power from 0 to 15 dBm. The data in Fig. \ref{QWP}d indicate that the QWP device is capable of responding to a high frequency modulated terahertz light with a carrier frequency around 4.2 THz. If we assume the rectified voltage has a frequency dependent roll off behavior of 1/[1+($\omega$$\tau$)$^2$], where $\omega$ is the microwave frequency and $\tau$ is a characteristic time including contributions from the intrinsic carrier relaxation time and the R-C circuit, the characteristics time $\tau$ at 0 dBm microwave power can be derived as $\tau$=1/(2$\pi$$f_{3\textrm{dB}}$)=37 ps. It has been known that for the intersubband devices the intrinsic carrier relaxation time is normally in few picoseconds level. Therefore, the current QWP device is working in the R-C dominant mode.

\begin{figure}[h]
\centering
\includegraphics[width=0.65\textwidth]{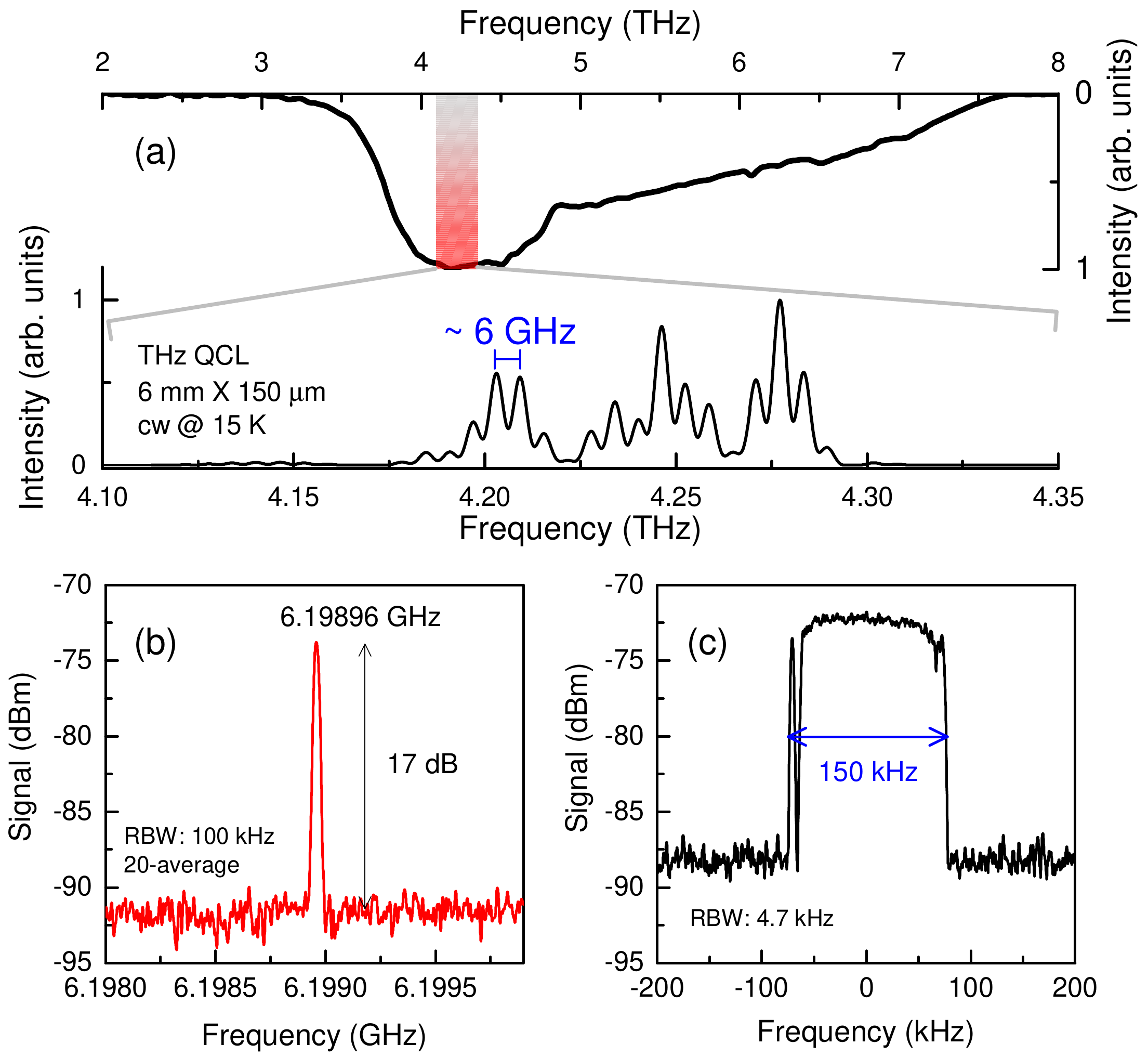}
\caption{\textbf{Fast detection of terahertz light.} (a) The emission spectrum of a 6-mm long terahertz QCL recorded in continuous wave mode at 15 K. The photoresponse spectrum of the terahertz QWP is shown for comparison. The reddish region shows the frequency range of the bottom X axis for the QCL measurement. (b) RF spectrum measured using the fast terahertz QWP. (c) RF spectrum recorded with the ``Max hold" function of the spectrum analyser. The time duration for the ``Max hold" measurement is 3 minutes. }
\label{BeatNote}
\end{figure}

In order to prove that the QWP device can work at high GHz frequency speed, we should find a suitable terahertz source for the fast detection. Actually the electrically-pumped terahertz QCL is an ideal source for this application due to the following reasons: (1) The emission frequency of a terahertz QCL can be intentionally designed to match the response frequency of the QWP, which therefore can result in strong photocurrent to improve the detection signal-to-noise ratio. (2) The terahertz light emitted from a terahertz QCL is naturally modulated at the cavity round trip frequency. Therefore by changing the cavity length, we are able to modify the laser modulation frequency and then evaluate the detection speed of the QWP device. As shown in Fig. \ref{QWP}d the response modulation bandwidth of the 400$\times$400 $\mu$m$^2$ QWP is measured to be 4.2 GHz at 0 dBm RF power. So we need a long cavity terahertz QCL to lase with a modulation frequency close to the response modulation bandwidth $\sim$4.2 GHz. Using a combined active region design of bound-to-continuum transition and resonant phonon scattering \cite{Giacomo2005,Wienold} for terahertz QCLs, we achieve an ultralow laser threshold current density of 50 A/cm$^2$ and thus the continuous wave (cw) operation of long cavity terahertz QCL is made possible \cite{WanSR}. Finally in this work we choose a 6-mm long terahertz QCL as a terahertz source for the fast detection. A typical emission spectrum of the QCL is shown in Fig. \ref{BeatNote}a (bottom X and left Y) measured using a Fourier transform infrared (FTIR) spectrometer with a spectral resolution of 3 GHz (0.1 cm$^{-1}$). As a reference, the response spectrum of the QWP is also plotted in Fig. \ref{BeatNote}a (top X and right Y). The reddish area (from 4.1 to 4.35 THz) represents the peak response range of the QWP. We can see that the QWP photoresponse spectrum exactly covers the QCL emission spectrum and the devices are perfectly spectral matched. The measured mode spacing of the terahertz QCL emission is 6 GHz which is roughly following $c/2ln$.

Using the experimental setup depicted in Fig. \ref{setup}, we successfully detect the fast modulated terahertz light. In Fig. \ref{BeatNote}b we show a typical RF spectrum measured with a spectrum analyser. The resolution bandwidth (RBW) used for this measurement is 100 kHz and the spectrum is obtained after 20 times average. A single line at 6.19896 GHz with a signal to noise ratio of 17 dB is clearly observed, which indicates that the QWP device can work as fast as 6.2 GHz although the rectification shows a response modulation bandwidth of 4.3 GHz. In frequency domain, we can say that the RF line shown in Fig. \ref{BeatNote} is originated from the inter-mode beating of the QCL longitudinal modes. Like the electrical beat note measurements \cite{GellieOE,LiOE}, the optical beat note signal presented in this work therefore can indirectly characterize the coherence properties of terahertz modes. In Fig. \ref{BeatNote}c we show the RF spectrum in ``Max hold" mode. With a time duration of 3 minutes, we see the spectrum spans over 150 kHz. The ``Max hold" linewidth can be regarded as an indicator of the laser frequency stability. It is worth noting that the optical beat note measurement is an alternative to the electrical beat note for accurately characterizing the repetition rate and mode stability of QCLs. Since the source and the detector are spatially separated, this space can be used for other applications; for instance, the optical beat note imaging \cite{BarbieriOE05}. Currently, the optical beat note signal obtained in this work is weak, i.e., -75 dBm power and 17 dB signal to noise ratio as shown in Fig. \ref{BeatNote}b. The weak signal would strongly prevent the technique from high resolution imaging applications. However, if the phase locking \cite{PL1,PL2} is implemented for the system, we can significantly improve the imaging resolution and contrast by employing the coherent imaging technique \cite{coherent-imaging}.

\section*{Discussion}

\begin{figure}[h]
\centering
\includegraphics[width=0.6\textwidth]{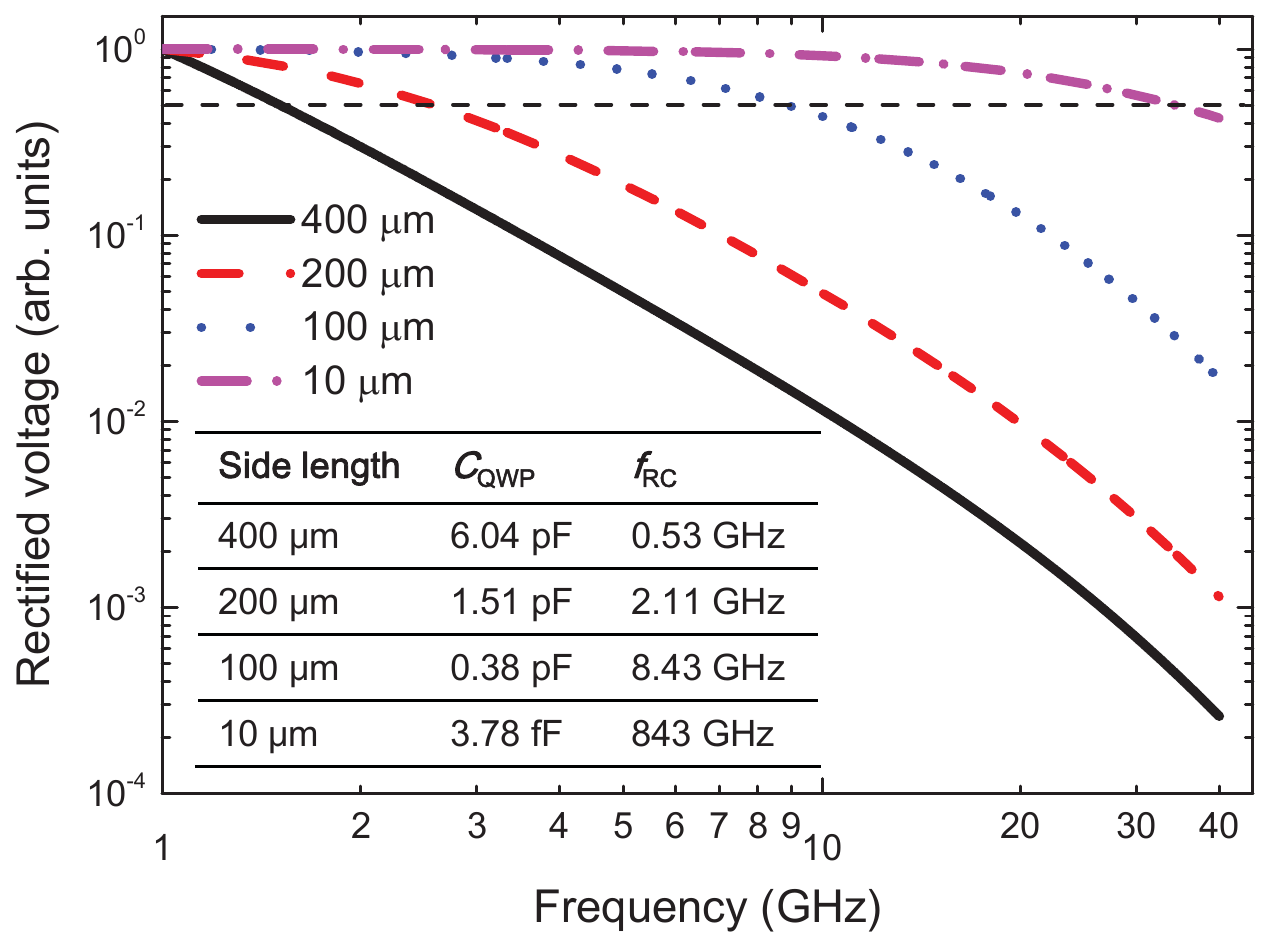}
\caption{\textbf{Modelling of rectified voltage for terahertz QWPs with different dimensions.} Calculated rectified voltage as a function of frequency upto 40 GHz for different mesa side lengths. For clear comparison, the rectified voltage is normalized.The horizontal dashed line shows the 3 dB attenuation power level. The inset gives the QWP capacitance $C_\textrm{QWP}$ calculated using a parallel-plate approximation and the R-C roll-off frequency $f_{\textrm{RC}}$=1/(2${\pi}R_\textrm{L}C_\textrm{QWP}$) for a load resistance $R_\textrm{L}$=50 $\Omega$ for different side lengths. }
\label{Cal}
\end{figure}

In this work we demonstrate that the fast terahertz QWP detector is capable of responding 6.2 GHz modulated terahertz light. We should remind that as we already elaborated in the text the 400$\times$400 $\mu$m$^2$ QWP device is working in the R-C dominant mode. Therefore, we have big space to further improve the response modulation bandwidth by optimizing the R-C circuit of the device. In Fig. \ref{Cal} we plot the calculated rectified voltage for different mesa side lengths using an expanded form of Equation \ref{rectV}
\begin{equation}\label{expan}
V_{\textrm{rect}}\propto|V^{\prime\prime}|_{0}\frac{1}{1+(\omega\tau_{\textrm{int}})^2}\frac{R_L}{(R_L+Z_{\textrm{dev}})^2}
\frac{1}{({\omega}C_{\textrm{QWP}}R_{\textrm{QWP}}-\textrm{j})^2},
\end{equation}
where j is the imaginary unit, $\tau_{\textrm{int}}$=4.8 ps \cite{LiuJQE1996} is the intrinsic carrier relaxation time, $Z_{\textrm{dev}}$=j$\omega{L}$+$R_{\textrm{QWP}}$/(1 +$\textrm{j}\omega{R_{\textrm{QWP}}}C_{\textrm{QWP}}$) is the total impedance of the QWP device with $L$ being the parasitic inductance caused by the short wire bonds, $R_{\textrm{QWP}}$ the differential resistance of the QWP obtained from the $V-I$ curve of Fig. \ref{QWP}a, $C_{\textrm{QWP}}$ the capacitance of the QWP. In this calculation, $R_{\textrm{QWP}}$ is set to 10$^5$ $\Omega$ level and $L$=0.1 nH. As shown in Fig. \ref{Cal}, with decreasing the side length to 10 $\mu$m, the 3 dB roll-off frequency is expected to reach 40 GHz. If we compare the 3 dB roll-off frequency with the R-C roll-off frequency, we can find that for larger side length, although the intrinsic carrier relaxation is fast, the roll-off is limited dominantly by the R-C circuit which leads to high device capacitance and low R-C roll-off frequency. However, once the side length is reduced to 10 $\mu$m, the situation is completely different. The calculated $f_\textrm{RC}$ for a 10 $\mu$m side length QWP is 843 GHz which is far greater than the overall 3 dB frequency of 40 GHz, which demonstrates that for small mesas the roll-off is limited dominantly by the intrinsic carrier relaxation time.

Due to the diffraction limit, the terahertz beam cannot be focused and confined in a 10$\times$10 $\mu$m$^2$ region. Therefore, to further improve the response speed of the current terahertz QWP, the device performance should be improved by, for instance, implementing grating coupler \cite{GratingQWP,MetalGrating} or patch antenna \cite{PalaferriAPL} techniques. On the other hand, the small QWP mesa can also work as a photomixer like a photoconductive antenna by using a metal-metal microcavity and a 50-$\Omega$ coplanar waveguide \cite{FPcavityMixer}.

It is worth pointing out that besides the terahertz QWP presented in this work, the Schottky barrier diodes (SBDs) and hot electron bolometers (HEBs) have been studied for long time and they also can be used either in direct detection or as nonlinear elements for frequency mixing (high speed operation) \cite{THzDet,Hussin,ItoJJAP,ZhangW,Shcheslavskiy}. For SBDs, the main advantages are that they are able to work at room temperature and the response frequency can reach 10 THz for direct detection. Although SBDs don't have the carrier relaxation time limitation for high speed operation, the intermediate frequency (IF) of SBD mixers is still limited by the device circuit due to the chip size, fabrication and packaging technology. Currently the commercial SBD mixers from Virginia Diodes can work up to few THz (greater than 2 THz is not commercially available) with an IF bandwidth of dozens of GHz \cite{VDI}. Note that these high speed fragile SBD mixers are highly customized and special attentions should be paid for operating them. Also a high local oscillator (LO) power greater than 1 mW is required for the SBD heterodyne detection \cite{Hubers}. Unlike SBDs, the superconducting HEB is a thermal detector only working at cryogenic temperature below 5 K. However, it is enough sensitive for single photon detection \cite{Shcheslavskiy}, while the NEP of SBDs lies in 10$^{-10}$ W/$\sqrt{\textrm{Hz}}$ level. In order to use a HEB as a mixer, it has to be fast. To achieve fast response for a desired IF of several GHz, the phonon-cooled microbridge should be thin enough (in few nanometers level) and the diffusion-cooled microbridge should be short (in few hundred nanometers level). The reported fastest HEB mixers based on NbN superconducting film can reach an IF bandwidth around 4-6 GHz \cite{Hubers}.

These two terahertz detectors, SBDs and HEBs, could be considered as strong competitors to terahertz QWPs. However, terahertz QWPs still show excellent advantages over SBDs and HEBs. Compared to SBDs, the terahertz QWP is more sensitive and the NEP is two orders of magnitude lower than that of the SBD. Therefore, in principle terahertz QWPs need less LO power for heterodyne detection than SBD mixers. In this work, we are working with the terahertz QWP at 4.2 THz which is far beyond the frequencies of the commercially available SBD mixers. Concerning the IF bandwidth, here we first demonstrate that the terahertz QWP can work as fast as 6.2 GHz. As shown in Fig. \ref{Cal}, the terahertz QWP is expected to work as fast as 40 GHz by optimizing the chip size. If the integrated coplanar waveguide is implemented, the speed of the terahertz QWP device can be further improved. We are hoping that after few years development the terahertz QWP can work as fast as, or even faster than the commercial room temperature SBD mixers. Compared to HEBs, terahertz QWPs show three main advantages. The first one is the operation temperature. Although terahertz QWP has to work at low temperatures, the cooling requirement is not that high as HEBs. We can operate terahertz QWPs at 20-30 K which is 20 K higher than HEBs. For the space heterodyne applications, this is a big improvement. The second one is that the HEB is easy to get saturated when it is illuminated by an external radiation, which makes it difficult for laser detection and active imaging applications. Nevertheless, the terahertz QWP has a much wider dynamic range and is suitable for various applications. The third advantage is the IF bandwidth. It is difficult to get an IF bandwidth larger than 7 GHz for HEB mixers. However, for terahertz QWPs, as we discussed before, the IF bandwidth can be significantly improved by optimizing chip size and antennas. A Single pixel HEB mixer pumped by a QCL LO has been already used for heterodyne detection of the neutral oxygen line \cite{Hayton}. Similarly, the terahertz QWP with a wavelength-matched QCL LO can be also used for terahertz heterodyne detections either on a balloon platform or in a satellite because of the above-mentioned advantages. Anyway, same as HEB mixers, it is still problematic to integrate the terahertz QWP and the QCL LO on a single chip though they are based on GaAs/AlGaAs materials and the fabrication processes for both devices are rather similar. One possibility to achieve the integration is to bond the QWP and QCL chips onto a patterned submount/carrier.

In conclusion we have demonstrated the fast detection of 6.2 GHz modulated terahertz light using a terahertz QWP equipped with a 50-$\Omega$ microwave strip line for high frequency signal extraction. The calculation of frequency-dependent rectified voltage has shown that the modulation response bandwidth is limited by the R-C circuit roll-off for larger size QWP mesa. However, for smaller mesa the response bandwidth is mainly limited by the intrinsic carrier relaxation. Besides the fast terahertz detection, the technique presented in this work can also be used for optically characterizing the frequency stability of terahertz QCLs, heterodyne detections and photomixing applications.

\section*{Methods}
\subsection*{Sample growth and device fabrication.}

The QWP is based on the one single photon design and the core region consists of 30-period AlGaAs/GaAs (80/18 nm) quantum well structure. The central 10 nm of the GaAs well is doped with Si to 5$\times$10$^{16}$ cm$^{-3}$. The core region is sandwiched between the top and bottom GaAs contact layers. The whole QWP structure is grown using a molecular beam epitaxy (MBE) system on a semi-insulating GaAs (100) substrate. The aluminum concentration of the AlGaAs barrier is of great importance for determining the peak response wavelength. Here the aluminum fraction is set to 1.9\%. The MBE-grown wafer is processed into mesa structures with top and bottom electrodes. The detailed fabrication process of the 45$^\circ$ edge facet QWP can be found in Ref. \cite{GuJPD}.

The QCL used in this work is based on a Al$_{0.25}$Ga$_{0.75}$As/GaAs material system grown by a MBE system on a semi-insulating GaAs (100) substrate. The growth starts with a 400-nm thick n-type GaAs bottom contact layer followed by the active region which consists of 76 cascade periods. Finally, a 50-nm thick GaAs top contact layer with 5$\times$10$^{18}$cm$^{-3}$ doping is grown on top of the active region. The MBE-grown QCL wafer is processed into single plasmon waveguide lasers using the standard semiconductor laser processing technology, for instance, the optical lithography, wet and dry etching, electron beam evaporation, lift-off, and thermal annealing.

The as-cleaved QWP and QCL chips are indium-soldered on the copper base. Wire bonds are then used for the electrical injection. Finally the packaged devices are screwed onto the cold fingers of continuous-flow liquid Helium cryostats for low temperature measurements.

\subsection*{Experimental characterizations.}
\textbf{Terahertz QWP performance characterization.} The photoresponse spectra of terahertz QWPs with a spectral resolution of 4 cm$^{-1}$ are measured using a Fourier Transform Infrared Spectrometer (FTIR) equipped with a Globar far-infrared broadband source. The peak responsivity of the QWP is calibrated using a 1000-K blackbody source. The noise spectral density is measured using a spectrum analyser with a resolution bandwidth of 1 Hz. To accurately measure the noise, a low noise current preamplifier (Stanford Research, SR570) with a sensitivity of 200 nA/V and a bandwidth of 2 kHz is used. The readout quantity from the spectrum analyser is in the unit of V/$\sqrt{\textrm{Hz}}$ which can be converted to A/$\sqrt{\textrm{Hz}}$ by considering the amplifier sensitivity in A/V. Finally the NEP parameter can be derived from the peak responsivity and the noise spectral density.

\textbf{Rectification measurement.} The microwave rectification of the terahertz QWP is measured by employing the experimental setup shown in Fig. \ref{QWP}c. To read the rectified voltage using a lock-in amplifier, we amplitude modulate the RF source at a frequency of 20 kHz by a wave function generator. In order to get a strong modulation signal, we increase the peak-to-peak voltage of the wave function generator until a 95\% modulation depth is obtained.

\textbf{Terahertz QCL characterization.} The emission spectrum of the 6-mm long terahertz QCL is measured using a FTIR. The terahertz light emitted from the QCL is coupled into the FTIR via a side input port using an off-axis parabolic mirror. The spectral resolution is set as 0.1 cm$^{-1}$. To reduce the water absorption, the FTIR is under vacuum and the beam path out side the FTIR is purged with dry air.

\textbf{Optical inter-mode beat note measurement.} The optical inter-mode beat note of the terahertz QCL is measured using the fast terahertz QWP as shown in Fig. \ref{setup}. The beat note spectra are recorded using a spectrum analyser with a bandwidth of 26 GHz. To measure the weak beat note signal, the optical alignment is critical. We do the alignment as follows: Firstly, we use a red diode laser with a piece of filter paper and a small aperture placed in sequence at the output port to imitate the QCL point source. It would be much easier to align the two parabolic mirrors using the visible analogous point source. After the optics are well aligned, we replace the diode laser with our QCL. Here we have to put the QCL facet exactly at the same position of the diode laser. Then, we run the QCL to do the fine alignment using a terahertz camera (NEC, IR/V-T0831C). Finally we replace the terahertz camera with the fast QWP to perform the optical beat note measurement.

\section*{Acknowledgements}

The authors thank Stefano Barbieri for discussions on fast terahertz detectors. This work was supported by the ``Hundred-Talent" Program of Chinese Academy of Sciences, the 973 Program of China (2014CB339803), the Major National Development Project of Scientific Instrument and Equipment (Grant No. 2011YQ150021), the National Natural Science Foundation of China (61575214 and 61405233), and the Shanghai Municipal Commission of Science and Technology (14530711300, 15560722000, 15ZR1447500, 15DZ0500103 and 17YF1430000).

\section*{Author contributions}

H.L. conceived the experiment, W.J.W. conducted the molecular beam epitaxy growth and fabricated the terahertz QCLs, Z.L.F. and H.X.W. fabricated the terahertz QWPs, Z.Y.T., Z.L.F., C.W. and H.L. performed the electrical, optical and microwave rectification characterizations for the terahertz QWP, H.L., W.J.W, and Z.P.L. conducted the optical inter-mode beat note measurements for the terahertz QCL using the fast terahertz QWP, H.L. modelled the frequency-dependent rectified voltage for terahertz QWPs, T.Z. debugged the LabView programs for the automated measurements and provided assistance during the measurements, X.G.G. designed the QWP structure, H.L. wrote the manuscript, H.L. and J.C.C. supervised the experiment, and all authors reviewed the manuscript.

\section*{Additional information}


\textbf{Competing financial interests:} The authors declear no competing financial interests.

\end{document}